\documentclass[prd,twocolumn,showpacs,amsmath,amssymb]{revtex4-2}
\usepackage{graphicx}
\usepackage{dcolumn}
\usepackage{bm}
\usepackage{rotating}
\usepackage{lipsum}
\usepackage{siunitx}
\usepackage{hhline}
\usepackage{multirow}
\usepackage{hyperref}
\usepackage[caption=false]{subfig}

\newcommand{\controlsearch}{\EE\to\eta\jpsi}
\newcommand{\controlnorm}{\EE\to\gamma_\text{ISR}\psi'}
\newcommand{\x}{X(3872)}
\newcommand{\xtopipijpsi}{\x\to\pip\pim\jpsi}
\newcommand{\xtopipizot}{\x\to\piz\piz\chi_{cJ}~(J=0,1,2)}
\newcommand{\xtopipiot}{\x\to\piz\piz\chicot}

\newcommand{\xtopipio}{\x\to\piz\piz\chico}
\newcommand{\xtopipit}{\x\to\piz\piz\chict}

\newcommand{\psip}{\psi^\prime}

\newcommand{\jpsi}{J/\psi}

\newcommand{\chicj}{\chi_{cJ}}
\newcommand{\chicz}{\chi_{c0}}
\newcommand{\chico}{\chi_{c1}}
\newcommand{\chict}{\chi_{c2}}
\newcommand{\chicot}{\chi_{c1,2}}
\newcommand{\EE}{e^+e^-}
\newcommand{\MM}{\mu^+\mu^-}
\newcommand{\pip}{\pi^+}
\newcommand{\pim}{\pi^-}
\newcommand{\piz}{\pi^0}

\newcommand{\bfg}{\begin{figure}}
\newcommand{\efg}{\end{figure}}
\newcommand{\bitm}{\begin{itemize}}
\newcommand{\eitm}{\end{itemize}}
\newcommand{\bnum}{\begin{enumerate}}
\newcommand{\enum}{\end{enumerate}}
\newcommand{\btbl}{\begin{table}}
\newcommand{\etbl}{\end{table}}
\newcommand{\btbu}{\begin{tabular}}
\newcommand{\etbu}{\end{tabular}}

\newcommand{\GG}{\gamma\gamma}

\newcommand{\LL}{l^+l^-}
\newcommand{\beq}{\begin{equation}}
\newcommand{\edq}{\end{equation}}
\newcommand{\g}{\gamma}
\newcommand{\bfchicjtogjpsi}{\mathcal{B}\left(\chicj\to\g\jpsi\right)}

\newcommand{\bfpizgg}{\mathcal{B}\left(\piz\to\GG\right)}
\newcommand{\bfnorm}{\mathcal{B}\left(\x\to\pip\pim\jpsi\right)}

\newcommand{\bfpiz}{\mathcal{B}\left(\x\to\piz\chicz\right)}
\newcommand{\bfpio}{\mathcal{B}\left(\x\to\piz\chico\right)}
\newcommand{\bfpit}{\mathcal{B}\left(\x\to\piz\chict\right)}
\newcommand{\bfpipiot}{\mathcal{B}\left(\xtopipiot\right)}
\newcommand{\bfpipij}{\mathcal{B}\left(\x\to\piz\piz\chicj\right)}

\newcommand{\bfpipio}{\mathcal{B}\left(\x\to\piz\piz\chico\right)}
\newcommand{\bfpipit}{\mathcal{B}\left(\x\to\piz\piz\chict\right)}

\newcommand{\rpiz}{\frac{\bfpiz}{\bfnorm}}
\newcommand{\rpio}{\frac{\bfpio}{\bfnorm}}
\newcommand{\rpit}{\frac{\bfpit}{\bfnorm}}
\newcommand{\rpipij}{\frac{\bfpipij}{\bfnorm}}

\newcommand{\chisqdof}{\chi^2/\text{d.o.f.}}
\newcommand{\nsig}{N_\text{sig}}
\newcommand{\nbkg}{N_\text{bkg}}
\newcommand{\nsearch}{N_\text{search}}

\newcommand{\sigregchico}{3.0}
\newcommand{\bkgregchico}{3.0}
\newcommand{\effchico}{7.9\%}
\newcommand{\tauchico}{0.184}
\newcommand{\yieldchico}{2.4^{~+1.8}_{~-1.9}}

\newcommand{\sigregchict}{0.0}
\newcommand{\bkgregchict}{6.0}
\newcommand{\effchict}{8.1\%}
\newcommand{\tauchict}{0.091}
\newcommand{\yieldchict}{0.0^{~+1.4}_{~-0.0}}

\newcommand{\sigregnorm}{107.0}
\newcommand{\bkgregnorm}{80.0}
\newcommand{\effnorm}{33.9\%}
\newcommand{\taunorm}{0.250}
\newcommand{\yieldnorm}{87.0^{~+10.9}_{~-10.3}}

\newcommand{\showersyso}{5.0}
\newcommand{\showersyst}{5.0}
\newcommand{\tracksyso}{2.0}
\newcommand{\tracksyst}{2.0}

\newcommand{\inputbfso}{2.9}
\newcommand{\inputbfst}{2.6}
\newcommand{\kinfitsyso}{1.7}
\newcommand{\kinfitsyst}{2.1}
\newcommand{\eopsyso}{2.7}
\newcommand{\eopsyst}{2.4}
\newcommand{\modelsyso}{11.0}
\newcommand{\modelsyst}{9.2}
\newcommand{\enerdepsyso}{0.4}
\newcommand{\enerdepsyst}{0.5}

\newcommand{\totalsyso}{13.0}
\newcommand{\totalsyst}{11.4}

\newcommand{\ulchico}{1.1}
\newcommand{\ulchict}{0.5}

\begin{document}
\normalsize
\parskip=5pt plus 1pt minus 1pt

\title{\boldmath Search for $\x\to\piz\piz\chi_{c1,2}$}
\author{
\begin{small}
\begin{center}
M.~Ablikim$^{1}$, M.~N.~Achasov$^{4,c}$, P.~Adlarson$^{75}$, O.~Afedulidis$^{3}$, X.~C.~Ai$^{80}$, R.~Aliberti$^{35}$, A.~Amoroso$^{74A,74C}$, Q.~An$^{71,58,a}$, Y.~Bai$^{57}$, O.~Bakina$^{36}$
% ...
\\
\vspace{0.2cm}
(BESIII Collaboration)\\
\vspace{0.2cm} {\it
$^{1}$ Institute of High Energy Physics, Beijing 100049, People's Republic of China\\
% ...
$^{3}$ Bochum  Ruhr-University, D-44780 Bochum, Germany\\
$^{4}$ Budker Institute of Nuclear Physics SB RAS (BINP), Novosibirsk 630090, Russia\\
% ...
$^{35}$ Johannes Gutenberg University of Mainz, Johann-Joachim-Becher-Weg 45, D-55099 Mainz, Germany\\
$^{36}$ Joint Institute for Nuclear Research, 141980 Dubna, Moscow region, Russia\\
% ...
$^{57}$ Southeast University, Nanjing 211100, People's Republic of China\\
$^{58}$ State Key Laboratory of Particle Detection and Electronics, Beijing 100049, Hefei 230026, People's Republic of China\\
% ...
$^{71}$ University of Science and Technology of China, Hefei 230026, People's Republic of China\\
% ...
$^{74}$ University of Turin and INFN, (A)University of Turin, I-10125, Turin, Italy; (B)University of Eastern Piedmont, I-15121, Alessandria, Italy; (C)INFN, I-10125, Turin, Italy\\
$^{75}$ Uppsala University, Box 516, SE-75120 Uppsala, Sweden\\
% ...
$^{80}$ Zhengzhou University, Zhengzhou 450001, People's Republic of China\\

\vspace{0.2cm}
$^{a}$ Deceased\\
% ...
$^{c}$ Also at the Novosibirsk State University, Novosibirsk, 630090, Russia\\
% ...

}

\end{center}
\vspace{0.4cm}
\end{small}
}

\date{September 19, 2024}

\begin{abstract}
Using 10.1~fb$^{-1}$ of $\EE$ collision data collected by the BESIII detector with center-of-mass energies between $\SI{4.15}{GeV}$ and $\SI{4.30}{GeV}$, we search for the decays $\xtopipiot$, where the $X(3872)$ is produced in $\EE\to\g\x$. No evidence above $3\sigma$ is found for either decay. Upper limits at the $90\%$ C.L. on the branching fractions of $\xtopipiot$ normalized to the branching fraction of $\xtopipijpsi$ are set to be 
$\bfpipio/\bfnorm < \ulchico$ and $\bfpipit/\bfnorm < \ulchict$, taking into account both statistical and systematic uncertainties. 
\end{abstract}


\maketitle

\section{Introduction}

In 2003, the Belle Collaboration reported the first observation of the state $X(3872)$, also referred to as $\chi_{c1}(3872)$, in the decay $B^\pm \to X(3872) K^\pm$~\cite{x3872belle}. This marked the beginning of the discovery of many charmonium-like states that exhibit clear discrepancies with a conventional charmonium ($c\bar{c}$) interpretation. While the $X(3872)$ has quantum numbers $J^{PC}=1^{++}$ allowed by the $c\bar{c}$ model, its mass differs from the nearest expected $c\bar{c}$ state, the $\chi_{c1}(2P)$, by $\SI{100}{MeV}/c^2$ \cite{charmoniumDifference}. The particularly narrow width of the $X(3872)$ of $\SI{1.19\pm0.21}{MeV}$ makes this difference especially stark~\cite{lhcbwidth1,lhcbwidth2}. In addition, the $X(3872)$ is known to decay through many isospin violating channels, including $\rho^0\jpsi$ \cite{x3872belle} and $\piz\chico$ \cite{ryan}.

Nearly two decades since its discovery, the $X(3872)$ lacks a definitive interpretation although several models have been proposed in order to explain its unusual properties. Due to its close proximity to the $D^0\bar{D}^{*0}$ threshold, the molecular $D^0\bar{D}^{*0} + \bar{D}^0D^{*0}$ interpretation offers a compelling possibility \cite{ddstarmolecule}. Two other explanations include a compact tetraquark interpretation \cite{compactTetraquark} or a superposition between a molecular and a conventional charmonium state \cite{superpositionState}. It is predicted that the branching fractions of the decays of the $X(3872)$ to $\pi\chicj$ or $\pi\pi\chicj (J=0,1,2)$ depend on the internal structure of the $X(3872)$ \cite{ccbar, molecular, updated_mehen}. These pionic transitions of the $X(3872)$ to $\chicj$ therefore serve as an excellent way to probe the quark configuration of the $\x$.

A previous BESIII analysis \cite{ryan} measured the single pion branching fraction ratio \[\rpio = 0.88^{~+0.33}_{~-0.27}~\pm0.10,\] where the first uncertainty is statistical and the second systematic, and placed upper limits of \[\rpiz < 19\] and \[\rpit < 1.1\] at the 90\% confidence level. Though these upper limits were consistent with both a conventional interpretation and a four-quark state (i.e. either a compact tetraquark or a hadronic molecule), the measurement on $\x\to\pi^0\chico$ was consistent with the four-quark state and disfavored the conventional picture. 
Another BESIII analysis \cite{will} reduced the upper limit on the $J=0$ decay to \[\rpiz < 3.6\] at the 90\% confidence level and placed upper limits on the two-pion decays to $\chi_{cJ}$. However, none of the aforementioned upper limits are stringent enough to rule out additional interpretations for the $X(3872)$.

In this paper, we strengthen these results by searching for the decays $\xtopipiot$ with BESIII data. For $J=1$, this decay is predicted to be extremely sensitive to the quark configuration of the $X(3872)$. For a conventional charmonium state, the ratio $\bfpipio/\bfpio$ is expected to be approximately 12.5 assuming the single $\piz$ is produced through two gluons \cite{ccbar}, although recent studies offer a critique of this prediction \cite{achasov_pi0, achasov_pi0pi0}. Meanwhile, for a four-quark state, the same ratio is predicted to be approximately $2.9\times10^{-3}$ \cite{updated_mehen}. This stark difference indicates that even an upper limit could provide key information disfavoring a conventional interpretation. Additionally, it is predicted that for $J=2$, this ratio is further suppressed and on the order of $10^{-6}$ \cite{molecular} for a molecular state.

We exclusively reconstruct the processes $\xtopipiot$, collectively called the search channel. The $\x$ is produced
alongside a photon, referred to as $\gamma_1$ throughout this paper, in $e^+e^-$ collisions. 
The $\chico$ and $\chict$ are reconstructed via $\chicot\to\g_2\jpsi$ and the $J/\psi$ is reconstructed via $J/\psi \to \EE$ or $J/\psi \to \MM$. We refer to the photon produced alongside the $\jpsi$ in $\chicot$ decays as $\g_2$.
The final state in the search channel is therefore $\g_1\g_2\piz\piz\LL$, where each $\piz$ is reconstructed from two photons and $\LL$ refers to either an electron or muon pair. 

With the same production of $\x$ in the decay $e^+e^-\to\g\x$, the branching fraction of the search channel is normalized to that of $\x\to\pip\pim\jpsi$, referred to as the normalization channel throughout this paper. This cancels common systematic uncertainties.

\section{Detector and data samples}

The BESIII detector~\cite{Ablikim:2009aa} records symmetric $e^+e^-$ collisions provided by the BEPCII storage ring~\cite{Yu:IPAC2016-TUYA01}, which operates with a peak luminosity of $1\times10^{33}$~cm$^{-2}$s$^{-1}$ in the center-of-mass (CM) energy range from 2.0 to 4.95~GeV. BESIII has collected large data samples in this energy region~\cite{Ablikim:2019hff}. The cylindrical core of the BESIII detector covers 93\% of the full solid angle and consists of a helium-based multilayer drift chamber~(MDC), a plastic scintillator time-of-flight system~(TOF), and a CsI(Tl) electromagnetic calorimeter~(EMC), which are all enclosed in a superconducting solenoidal magnet providing a 1.0~T magnetic field. The solenoid is supported by an octagonal flux-return yoke with resistive plate counter muon identification modules interleaved with steel. 
The charged-particle momentum resolution at $1~{\textrm{GeV}}/c$ is $0.5\%$, and the  ${\textrm d}E/{\textrm d}x$ resolution is $6\%$ for electrons from Bhabha scattering. The EMC measures photon energies with a resolution of $2.5\%$ ($5\%$) at $1$~GeV in the barrel (end cap) region. The time resolution in the TOF barrel region is 68~ps, while that in the end cap region is 110~ps. The end cap TOF system was upgraded in 2015 using multigap resistive plate chamber technology, providing a time resolution of 60~ps~\cite{etof1,etof2,etof3}.

For this analysis, we study the BESIII data collected at center-of-mass energies between 4.15~GeV and 4.30~GeV corresponding to an integrated luminosity of 10.1~fb$^{-1}$. In this range, the cross section $\sigma\left[(\EE\to\g\x\right]$ has been found to be the largest \cite{eetogx1,eetogx2}. The full list of data points used is provided in Table \ref{tab:data} and Table \ref{tab:rscan}. 

\begin{table}[!htbp]
    \centering
    \caption{\label{tab:data} Data samples with center-of-mass energies between \SI{4.15}{GeV} and \SI{4.30}{GeV}. Points without references are preliminary.}
    \begin{tabular}{cc}
        \hhline{==}
        Center-of-Mass Energy (MeV) & Luminosity (pb\textsuperscript{-1}) \\
        \hline
        $4157.83\pm0.05\pm0.36$ \cite{lumecm2017}   & $\phantom{0}411.31\pm2.55$ \phantom{[00]} \\ 
        $4178\phantom{.00\pm0.00\pm0.00}$ \phantom{[00]}   & $3194.5\phantom{0}\pm0.2\phantom{0}$ \phantom{[00]} \\
        $4188.59\pm0.15\pm0.68$ \cite{ecm2010-2014} & $\phantom{00}43.09\pm0.03$ \cite{lum2010-2014} \\ 
        $4189.12\pm0.05\pm0.34$ \cite{lumecm2017}   & $\phantom{0}526.70\pm2.16$ \cite{lumecm2017}   \\
        $4199.15\pm0.05\pm0.34$ \cite{lumecm2017}   & $\phantom{0}526.60\pm2.05$ \cite{lumecm2017}   \\
        $4207.73\pm0.14\pm0.61$ \cite{ecm2010-2014} & $\phantom{00}54.55\pm0.03$ \cite{lum2010-2014} \\ 
        $4209.39\pm0.06\pm0.34$ \cite{lumecm2017}   & $\phantom{0}517.10\pm1.81$ \cite{lumecm2017}   \\
        $4217.13\pm0.14\pm0.67$ \cite{ecm2010-2014} & $\phantom{00}54.13\pm0.03$ \cite{lum2010-2014} \\
        $4218.93\pm0.06\pm0.32$ \cite{lumecm2017}   & $\phantom{0}514.60\pm1.80$ \cite{lumecm2017}   \\
        $4226.26\pm0.04\pm0.65$ \cite{ecm2010-2014} & $\phantom{00}44.40\pm0.03$ \cite{lum2010-2014} \\
        $4226.26\pm0.04\pm0.65$ \cite{ecm2010-2014} & $\phantom{}1047.34\pm0.14$ \cite{lum2010-2014} \\
        $4235.77\pm0.04\pm0.30$ \cite{lumecm2017}   & $\phantom{0}530.30\pm2.39$ \cite{lumecm2017}   \\
        $4241.66\pm0.12\pm0.73$ \cite{ecm2010-2014} & $\phantom{00}55.59\pm0.04$ \cite{lum2010-2014} \\
        $4243.97\pm0.04\pm0.30$ \cite{lumecm2017}   & $\phantom{0}538.10\pm2.69$ \cite{lumecm2017}   \\
        $4257.97\pm0.04\pm0.66$ \cite{ecm2010-2014} & $\phantom{0}523.74\pm0.10$ \cite{lum2010-2014} \\
        $4257.97\pm0.04\pm0.66$ \cite{ecm2010-2014} & $\phantom{0}301.93\pm0.08$ \cite{lum2010-2014} \\  
        $4266.81\pm0.04\pm0.32$ \cite{lumecm2017}   & $\phantom{0}531.10\pm3.13$ \cite{lumecm2017}   \\
        $4277.78\pm0.11\pm0.52$ \cite{lumecm2017}   & $\phantom{0}175.70\pm0.97$ \cite{lumecm2017}   \\   
        $4288.43\pm0.06\pm0.34$ \cite{lumecm2017}   & $\phantom{0}501.18\pm3.11$ \phantom{[00]} \\    
        \hline
        29 energies at low luminosities & See Table \ref{tab:rscan} \\ 
        \hhline{==}
    \end{tabular}
\end{table}
\begin{table}[!h]
    \centering
    \caption{\label{tab:rscan} Low-luminosity scan data samples with center-of-mass energies between \SI{4.15}{GeV} and \SI{4.30}{GeV}. The first error on each luminosity is statistical and the second is systematic.}
    \begin{tabular}{cc}
        \hhline{==}
        Center-of-mass energy (MeV) & Luminosity (pb\textsuperscript{-1}) \\
        \hline
        $4150$ & $7.662 \pm 0.018 \pm 0.053$ \cite{rscan_lum} \\
        $4160$ & $7.954 \pm 0.019 \pm 0.056$ \cite{rscan_lum} \\
        $4170$ & $8.008 \pm 0.039 \pm 0.130$ \cite{rscan_lum} \\
        $4180$ & $7.309 \pm 0.018 \pm 0.051$ \cite{rscan_lum} \\
        $4190$ & $7.560 \pm 0.018 \pm 0.052$ \cite{rscan_lum} \\
        $4195$ & $7.503 \pm 0.018 \pm 0.054$ \cite{rscan_lum} \\
        $4200$ & $7.582 \pm 0.018 \pm 0.053$ \cite{rscan_lum} \\
        $4203$ & $6.815 \pm 0.017 \pm 0.048$ \cite{rscan_lum} \\
        $4206$ & $7.638 \pm 0.018 \pm 0.055$ \cite{rscan_lum} \\
        $4210$ & $7.678 \pm 0.018 \pm 0.054$ \cite{rscan_lum} \\
        $4215$ & $7.768 \pm 0.019 \pm 0.054$ \cite{rscan_lum} \\
        $4220$ & $7.935 \pm 0.019 \pm 0.055$ \cite{rscan_lum} \\
        $4225$ & $8.212 \pm 0.020 \pm 0.061$ \cite{rscan_lum} \\
        $4230$ & $8.193 \pm 0.020 \pm 0.057$ \cite{rscan_lum} \\
        $4235$ & $8.273 \pm 0.020 \pm 0.057$ \cite{rscan_lum} \\
        $4240$ & $7.830 \pm 0.019 \pm 0.054$ \cite{rscan_lum} \\
        $4243$ & $8.571 \pm 0.020 \pm 0.060$ \cite{rscan_lum} \\
        $4245$ & $8.487 \pm 0.020 \pm 0.060$ \cite{rscan_lum} \\
        $4248$ & $8.554 \pm 0.020 \pm 0.059$ \cite{rscan_lum} \\
        $4250$ & $8.596 \pm 0.020 \pm 0.060$ \cite{rscan_lum} \\
        $4255$ & $8.657 \pm 0.020 \pm 0.060$ \cite{rscan_lum} \\
        $4260$ & $8.880 \pm 0.021 \pm 0.063$ \cite{rscan_lum} \\
        $4265$ & $8.629 \pm 0.020 \pm 0.061$ \cite{rscan_lum} \\
        $4270$ & $8.548 \pm 0.020 \pm 0.060$ \cite{rscan_lum} \\
        $4275$ & $8.567 \pm 0.020 \pm 0.060$ \cite{rscan_lum} \\
        $4280$ & $8.723 \pm 0.021 \pm 0.060$ \cite{rscan_lum} \\
        $4285$ & $8.596 \pm 0.020 \pm 0.059$ \cite{rscan_lum} \\
        $4290$ & $9.010 \pm 0.021 \pm 0.062$ \cite{rscan_lum} \\
        $4300$ & $8.453 \pm 0.020 \pm 0.064$ \cite{rscan_lum} \\
        \hhline{==}
    \end{tabular}
\end{table}

Monte Carlo (MC) simulated data samples generated with a {\sc geant4}-based~\cite{geant4} software package, which includes the geometric description of the BESIII detector and the detector response, are used to determine detection efficiencies and to estimate backgrounds. The simulation models the beam energy spread and initial state radiation (ISR) in the $e^+e^-$ annihilations with the generator {\sc kkmc}~\cite{kkmc1,kkmc2}. The inclusive MC sample includes the production of open charm processes, the ISR production of vector charmonium(-like) states, and the continuum processes incorporated in {\sc kkmc}~\cite{kkmc1,kkmc2}. All particle decays are modeled with {\sc evtgen}~\cite{evtgen1,evtgen2} using branching fractions  either taken from the Particle Data Group (PDG)~\cite{pdg}, when available, or otherwise estimated with {\sc lundcharm}~\cite{lundcharm1,lundcharm2}. Final state radiation from charged final-state particles is incorporated using the {\sc photos} package~\cite{photos}.

Signal MC samples are generated for the processes $\EE\to\g\x$ with $\xtopipizot$ and $\xtopipijpsi$ for the purposes of calculating the reconstruction efficiency and optimizing the event selection. The decay $\EE\to\g\x$ is taken to be dominated by an electric dipole transition with the angular distribution described as $1-\tfrac{1}{3}\cos^{2}\theta$. The decays of the $\x$ to $\piz\piz\chicj$ are generated uniformly according to a phase space model. The normalization channel decays $\xtopipijpsi$ are described with an $S$-wave model through the intermediate process $\x\to\rho^0\jpsi$ with $\rho\to\pip\pim$. All resonance decays assume a Breit-Wigner mass dependence. For visualization, all three search channel modes are generated with the same branching fraction $\bfpio$ measured in Ref. \cite{ryan}. Samples are generated with four thousand times more events than expected in the data and scaled down to reduce statistical fluctuations. 

A variety of exclusive background MC samples are generated to study $\jpsi$ peaking backgrounds. 
These backgrounds share final states identical to, or potentially mis-identified as, the search channel, resulting in peaking backgrounds in or near the signal region. The number of events generated for each decay is computed as a product of previously measured luminosities, cross sections \cite{pdg, xs_x3872, xs_x3823, xs_omegachic0, xs_omegachic2, xs_pipipsip, xs_etajpsi, xs_etapjpsi}, and branching fractions. Samples are generated with excess statistics and scaled down to the appropriate luminosity. The $\jpsi$ decays only to $\EE$ or $\MM$, and the $\psi(3686)$ may decay to $\pi\pi\jpsi$, $\eta\jpsi$, $\piz\jpsi$, $\g\g\jpsi$, or $\g\chi_{cJ}$. The background decays studied for this analysis are listed in Table \ref{tab:excmc}. All light hadrons decay inclusively.

\begin{table}[!htbp]
    \caption{\label{tab:excmc} Decays used to generate exclusive MC. }

    \begin{center}
        \begin{tabular}{cc}
            \hhline{==}
            Group & Process\\
            \hline
            \multirow{7}{*}{$J/\psi$ peaking} & $e^+e^-\rightarrow\omega\chi_{cJ}$ \\
                                                         & $e^+e^-\rightarrow\eta J/\psi$ \\
                                                         & $e^+e^-\rightarrow\eta' J/\psi$ \\
                                                         & $e^+e^-\rightarrow\pi^+\pi^- \psi'$ \\
                                                         & $e^+e^-\rightarrow\pi^0\pi^0 \psi'$ \\
                                                         & $e^+e^-\rightarrow\pi^+\pi^- J/\psi$\\
                                                         & $e^+e^-\rightarrow\pi^0\pi^0 J/\psi$\\
            \hline
            X(3823) peaking                              & $e^+e^-\rightarrow\pi^0\pi^0X(3823)$ \\
            \hline
            ISR & $e^+e^- \rightarrow \gamma \psi'$ \\
            \hline
            \multirow{5}{*}{$X(3872)$ peaking} & $X(3872) \rightarrow \pi^+\pi^-J/\psi$ \\
                                                          & $X(3872) \rightarrow \pi^0\chi_{cJ}$ \\
                                                          & $X(3872) \rightarrow \gamma \psi'$ \\
                                                          & $X(3872) \rightarrow \gamma J/\psi$\\
                                                          & $X(3872) \rightarrow \omega J/\psi$\\
            \hhline{==}
        \end{tabular}
    \end{center}

\end{table}

The $\jpsi$ sideband events are used to estimate non-$\jpsi$ backgrounds. 
We choose the $\jpsi$ sideband using $\SI{44}{MeV}/c^2 \leq |M(\LL)-M(\jpsi)| \leq \SI{176}{MeV}/c^2$, where $M(\LL)$ is the measured invariant mass of the final-state leptons and $M(\jpsi)$ is the nominal $\jpsi$ mass tabulated in the PDG \cite{pdg}, resulting in a sideband four times wider than the signal region.

\section{Event Selection}

\label{sec:event_selection}

The final states for both the normalization and search channels are fully reconstructed. 
For the normalization channel, we use the same event selection criteria as in the single-pion analysis \cite{ryan}. Details for the selection of the search channel are described in the following.

Charged tracks detected in the MDC are required to be within a polar angle ($\theta$) range of $|\rm{cos\theta}|<0.93$, where $\theta$ is defined with respect to the $z$-axis, which is the symmetry axis of the MDC. Their distance of closest approach to the interaction point (IP) must be less than 10\,cm along the $z$-axis, $|V_{z}|$, and less than 1\,cm in the transverse plane, $|V_{xy}|$. 
To distinguish between electrons and muons, each electron pair is required to have at least one candidate with $E/p>0.85$, while each muon pair requires both candidates to have $E/p<0.25$, where $E$ is the energy of the candidate deposited in the EMC and $p$ is the momentum of the candidate reconstructed in the MDC.
To select $\jpsi$ events, we require the $\LL$ mass to be within $\SI{33}{MeV}/c^2$ of the nominal $\jpsi$ mass.

Photons are identified as showers within the EMC. Energy deposited in nearby TOF counters is added to the shower energy. The deposited energy of each shower must be more than 25~MeV in the barrel region ($|\cos \theta|< 0.80$) and more than 50~MeV in the end cap region ($0.86 <|\cos \theta|< 0.92$). To suppress electronic noise and showers unrelated to the event, the difference between the EMC time and the event start time is required to be within 
[0, 700]\,ns. 
The $\piz$ candidates are formed from pairs of photons and are required to satisfy $\SI{107}{MeV}/c^2 < M(\g\g) < \SI{163}{MeV}/c^2$.  In addition, a $\pi^0$ mass fit is performed separately for each pair and $\chi^2<2500$ is imposed.

A six constraint (6C) kinematic fit is performed on the final-state particles. Momentum and energy account for four of these constraints, while the remaining two constraints comprise $\pi^0$ mass constraints. 
We require $\chi_{6C}^2/\text{d.o.f}<9.0$ for $\chico$ events and $\chi_{6C}^2/\text{d.o.f}<8.0$ for $\chict$ events, where $\chi_{6C}^2/\text{d.o.f}$ is the 6C kinematic fit $\chi^2$ statistic per degree of freedom. These values were chosen by optimizing over a figure of merit~(FOM)
\begin{equation}
	\label{eq:fom}
    FOM = \sqrt{-2 \ln{\frac{P(\mu \tau; N_\text{bkg}) P(\mu; N_\text{sig})}{P(N_\text{bkg};N_\text{bkg})P(N_\text{sig};N_\text{sig})}}},
\end{equation}
which is a direct computation of the expected significance where $P$ is the Poisson distribution, $N_\text{sig}$ is the number of events in the signal region, $N_\text{bkg}$ is the number of events in the background region, and $\tau$ is the expected ratio of background events in the background region to background events in the signal region. The signal and background regions are described in the next section.  The free parameter $\mu$ is chosen to maximize the numerator. Because $N_\text{sig}$ and $N_\text{bkg}$ are not integers in exclusive MC, the figure of merit is computed by sampling Poisson distributions with means $N_\text{sig}$ and $N_\text{bkg}$ and averaging the results. 

\begin{figure*}[t]
    \hspace*{\fill}
    \subfloat[][]{\label{sfig:plots2d_dat}\includegraphics[width=0.3\textwidth]{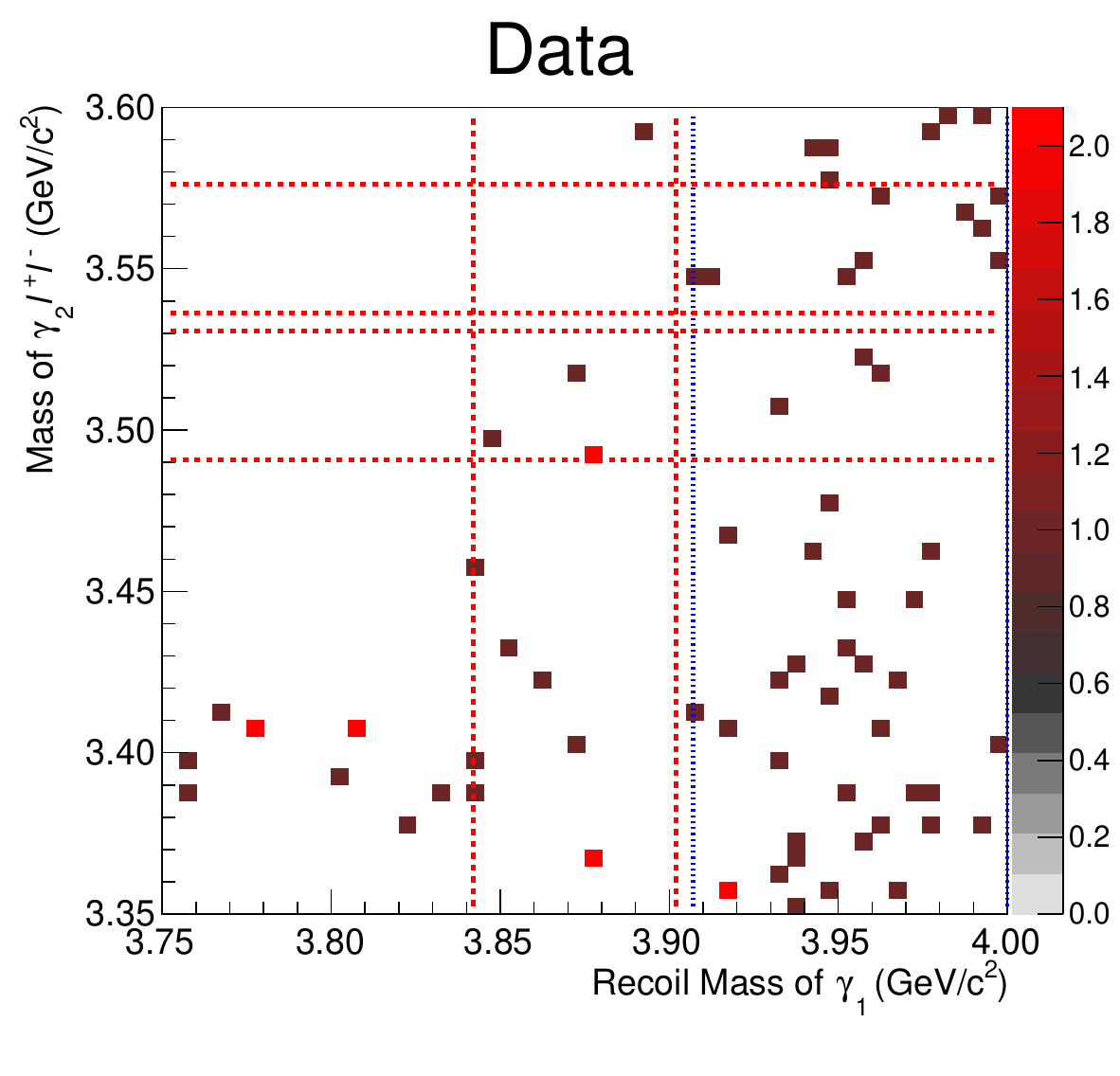}}
    \hfill
    \subfloat[][]{\label{sfig:plots2d_sig}\includegraphics[width=0.3\textwidth]{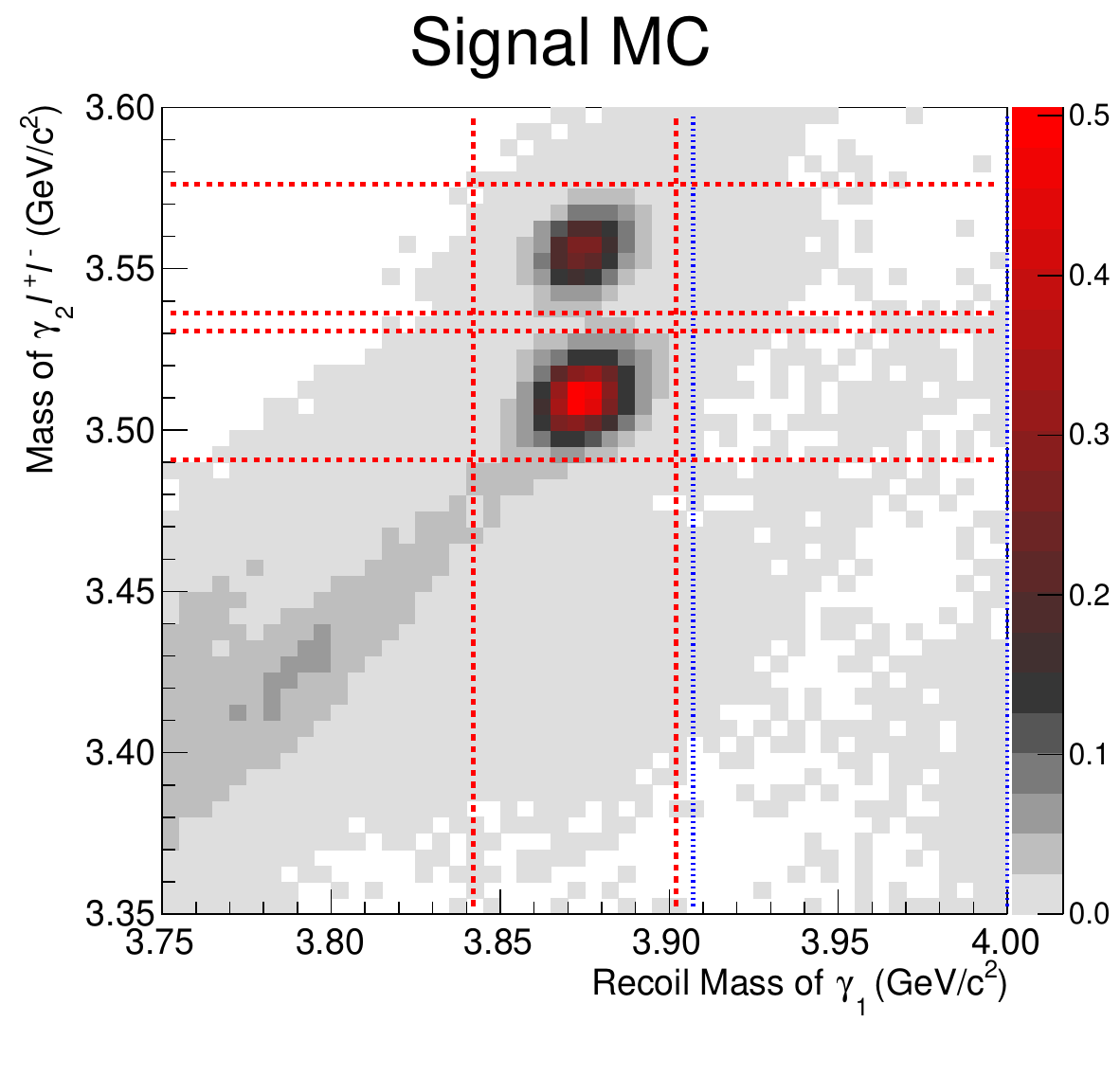}}
    \hfill
    \subfloat[][]{\label{sfig:plots2d_bkg}\includegraphics[width=0.3\textwidth]{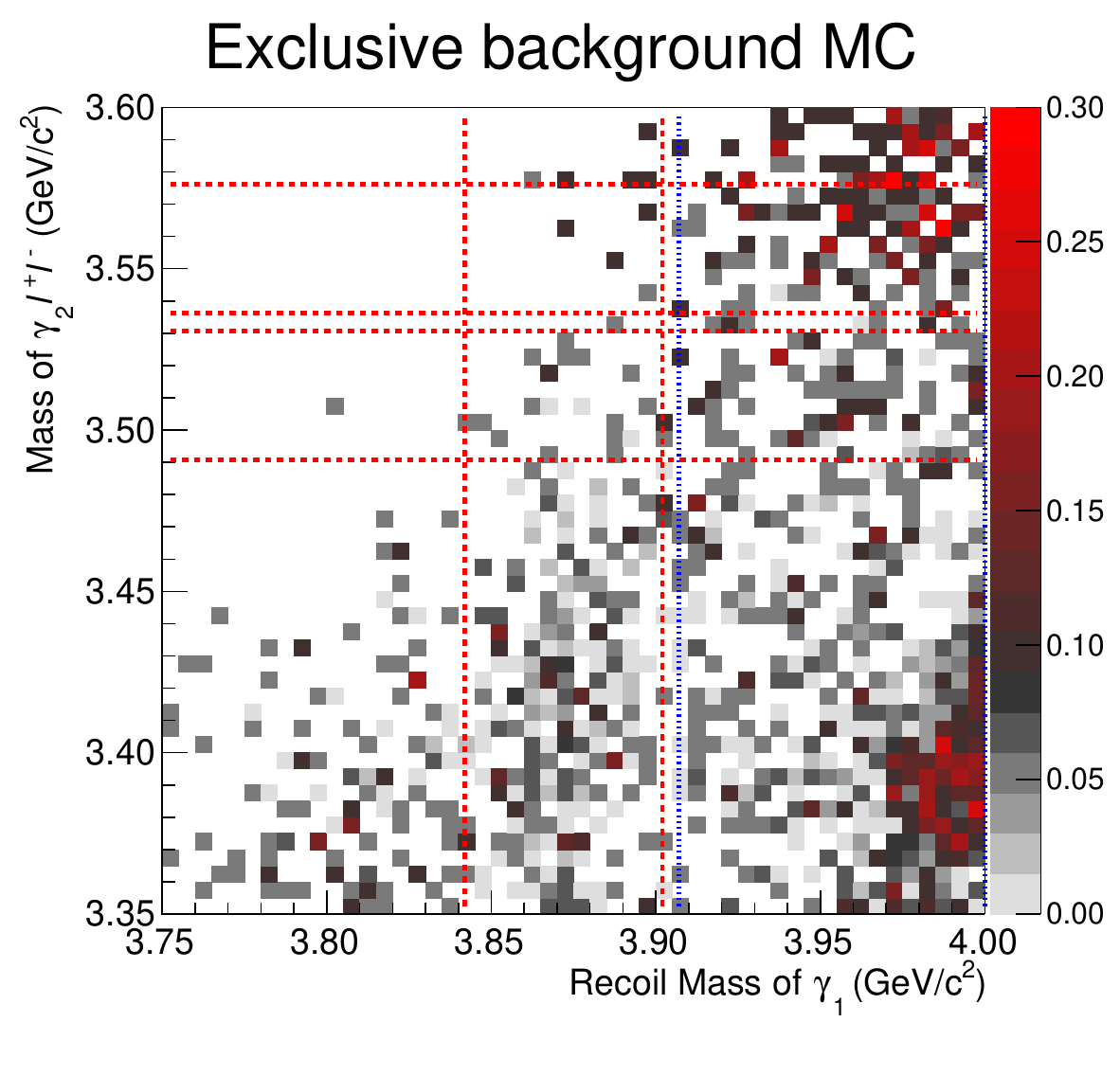}}
    \hspace*{\fill}
    \caption{\label{fig:plots2d} Invariant mass spectrum for $\gamma_2\LL$ versus the recoil mass of $\gamma_1$ around the signal region for data (left), signal MC (center), and exclusive background MC (right). Events are subject to all selection criteria discussed in Section \ref{sec:event_selection} except those eliminating multiple combinations. Dashed red lines denote the $\x$ signal region in the search channel and the dotted blue lines indicate the sideband region in the search channel. Due to limited phase space below the $\x$ mass, only the high mass sideband is used. Though present in the MC, no $\chicz$ signal is seen. An excess of events in the lower-left of the signal MC plot is the result of incorrectly combined $\chicot$ candidates owing to the presence of two photons in the search channel final state.}
\end{figure*}

Various vetoes are also applied to the search channel to reject the peaking backgrounds modeled by the exclusive background MC listed in Table \ref{tab:excmc} from the signal region. These include an $\eta^{(\prime)}\to\g\g\piz\piz$ veto to remove $\EE\to\eta^{(\prime)}\jpsi$ and a $\psi(3686)\to\g\g\LL$ veto to remove $\EE\to\piz\piz\psi(3686)$.
These vetoes are optimized by the FOM defined in Eq. \ref{eq:fom}. A list of all optimized requirements is given in Table \ref{tab:optimized_cuts}.

In addition, due to the large number of photons in the final state, we address the issue of multiple-counting through two approaches. For photons originating from $\piz$ decays, we select the combination with the lowest $\chisqdof$ value from the kinematic fit. The remaining two photons, not associated with $\piz$ decays, are differentiated by selecting the mass combination of any photon with $\gamma_2\LL$ closest to the relevant world average $\chi_{cJ}$ mass \cite{pdg}.

After these selection criteria, Fig.~\ref{fig:plots2d} shows the invariant mass of the $\gamma_2 \LL$ system vs. the mass recoiling against $\gamma_1$. The signal MC sample shows
clear $\chico$ and $\chict$ signals, but no $\chicz$ due to its strongly suppressed radiative decay.

\begin{figure*}
	\centering
	\hspace*{\fill}
    \includegraphics[width=0.4\textwidth]{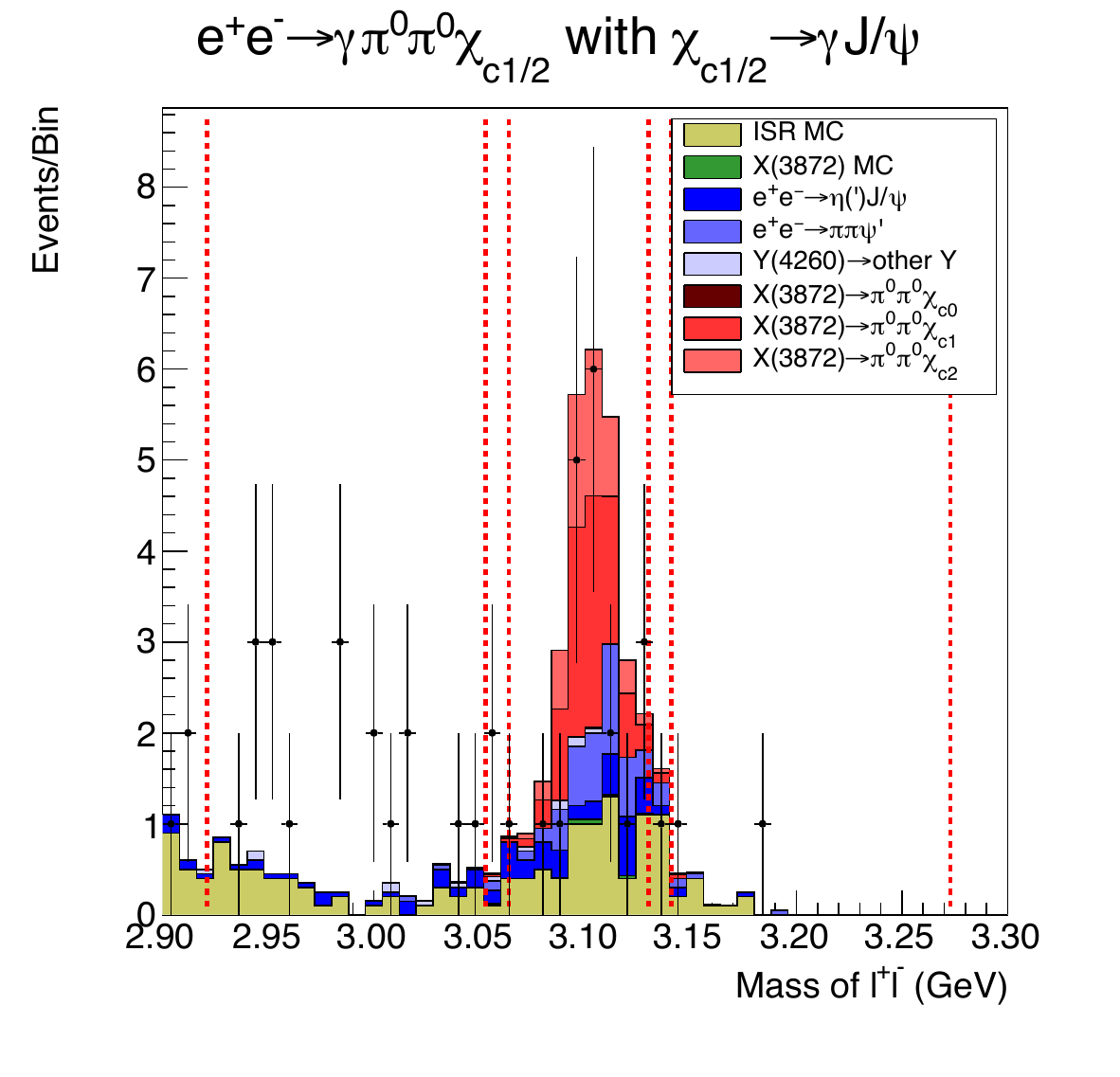}
    \hfill
    \includegraphics[width=0.4\textwidth]{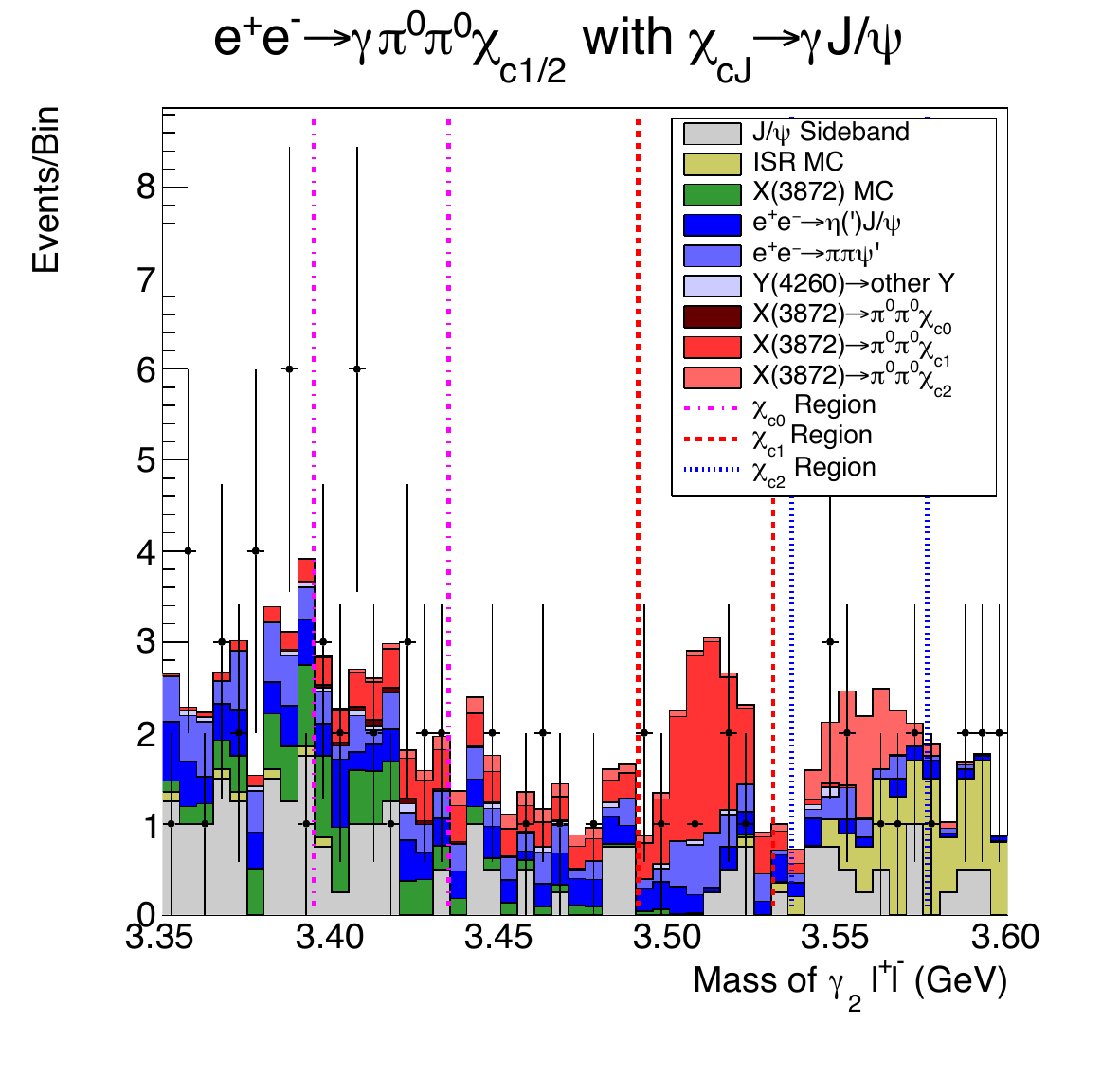}
    \hspace*{\fill}
    \caption{\label{fig:alternate_projections} Masses of $\jpsi$ (left) and $\chicj$ (right) candidates. The stacked histograms give the MC and, on the right, the $\jpsi$ data sidebands. The histogram labeled ``other Y'' contains all $\jpsi$-peaking background MC not specifically plotted with its own color. The points with error bars are the data. On the left, the center lines denote the $\jpsi$ signal region in the search channel and the outer lines denote the boundaries of the $\jpsi$ data sidebands. On the right, the leftmost pair of lines indicates the boundaries of the $\chicz$ signal region, which we are not sensitive to, while the center and right pairs show the $\chico$ and $\chict$ signal selections, respectively. All cuts are included except those on the variable along the horizontal axis and the minimum $\chisqdof$ selection. For the right plot, the $\chisqdof$ requirement for $\chico$ events is used for all $\chicj$ candidate masses. }
\end{figure*}

To select $\chicj$ candidates, we require $M(\gamma_2\LL)-M(\LL)+M(\jpsi)$ to be within $\SI{20}{MeV}$ of the nominal $\chicj$ mass for a particular $J$. 
The invariant mass spectra of the $\LL$ and $\gamma_2\LL$ systems, along with their corresponding requirements, are shown in Fig. \ref{fig:alternate_projections}.

\begin{table}[!htbp]
    \caption{\label{tab:optimized_cuts} Processes we choose to veto and their optimized selection criteria. }
    \begin{center}
        \begin{tabular}{cc}
            \hhline{==}
            Cut & Value\\
            \hline
            Kinematic fit ($J=1$) & $\chi^2/\text{d.o.f.} < 9.0$\\
            Kinematic fit ($J=2$) & $\chi^2/\text{d.o.f.} < 8.0$\\
            \hline
            $\EE\to\eta\jpsi$ veto & $|M(\g_1\g_2\piz\piz) - M(\eta)| > \SI{40}{MeV}/c^2$\\
            $\EE\to\eta'\jpsi$ veto & $|M(\g_1\g_2\piz\piz) - M(\eta')| > \SI{15}{MeV}/c^2$\\
            $\EE\to\piz\piz\psip$ veto & $|M(\g_1\g_2\LL) - M(\psip)| > \SI{30}{MeV}/c^2$\\
            \hhline{==}
        \end{tabular}
    \end{center}

\end{table}

\section{Signal Yields}

\label{sec:yields}

\begin{figure*}
	\centering
	\hspace*{\fill}
        \includegraphics[width=0.4\textwidth]{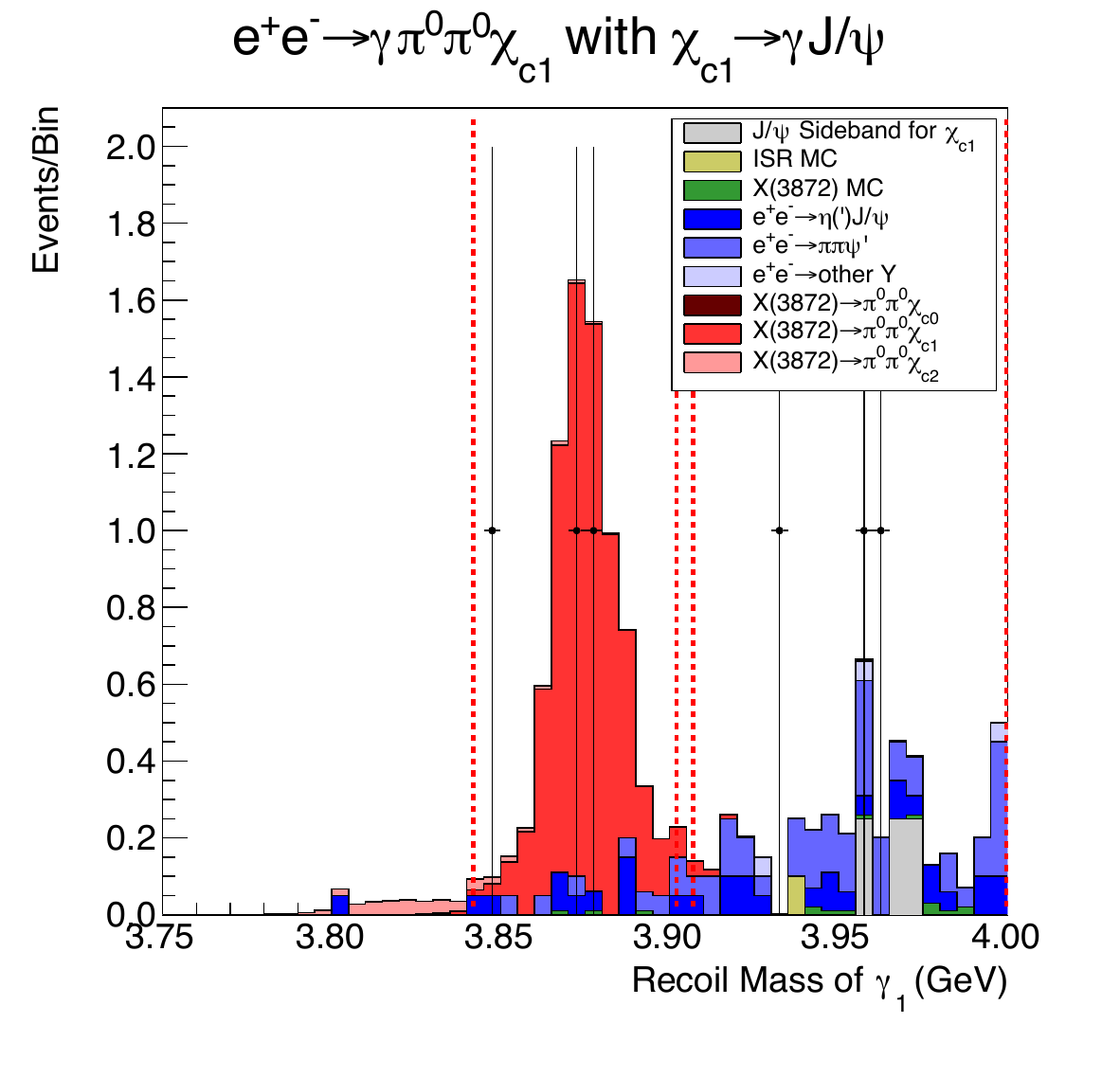}
    \hfill
        \includegraphics[width=0.4\textwidth]{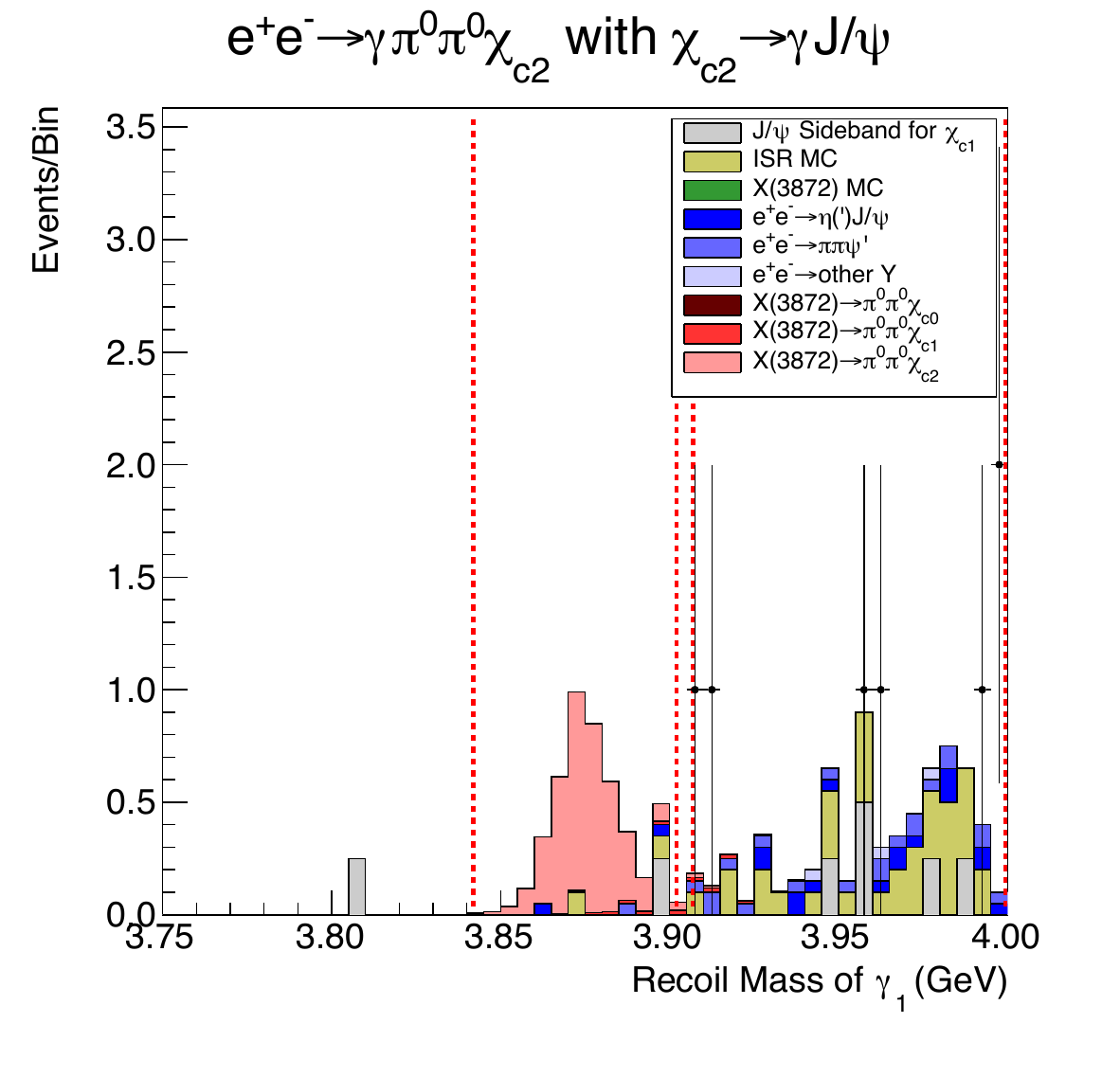}
    \hspace*{\fill}
    \caption{\label{fig:search_yields} Masses of $\x$ candidates in the $\chico$ and $\chict$ regions. The data is given by points, while the stacked histograms represent exclusive background MC and $\jpsi$ data sidebands. The histogram labeled ``other Y'' contains all $\jpsi$-peaking background MC not specifically plotted with its own color. The two leftmost dashed lines give the signal region and the two rightmost dashed lines give the background regions. A counting method is used to determine the number of $\x$ events in the signal region. }
\end{figure*}

Due to the low number of events in the search channel, signal yields are determined using a counting method. We measure the number of events $N_\text{sig}$ in the signal region, defined as the region in the recoil mass spectrum of $\gamma_1$ against the CM within $\SI{30}{MeV}$ of the nominal $\x$ mass, and the number of events $N_\text{bkg}$ in the background region, defined as the region $\SI{35}{MeV}$ above the nominal $\x$ mass in the recoil mass spectrum of $\gamma_1$ against the CM and extending to the end of the plot window at $\SI{4.0}{GeV}$. Phase space is limited at masses below the $\x$ peak, so this region is not considered to estimate the background. 
There is good agreement between the number of data events in the $X(3872)$ sidebands and the number of events predicted by the $\jpsi$ sidebands and exclusive background MC in those same $X(3872)$ sidebands. 
As such, we estimate the background in the signal region by assuming the ratio $\tau$ of background events in the signal region to the number of background events in the sideband region is the same as the ratio calculated from MC. The signal yield in the search channel is therefore  $\nsearch = \nsig - \tau\nbkg$. Asymmetric $1\sigma$ uncertainties on the yield are calculated using $\nsig$, $\nbkg$, and $\tau$ as inputs for the Rolke method \cite{rolke}. Data and MC measured in each region of the search channel are plotted in Fig. \ref{fig:search_yields}.

\begin{figure}
	\centering
    \includegraphics[width=0.4\textwidth]{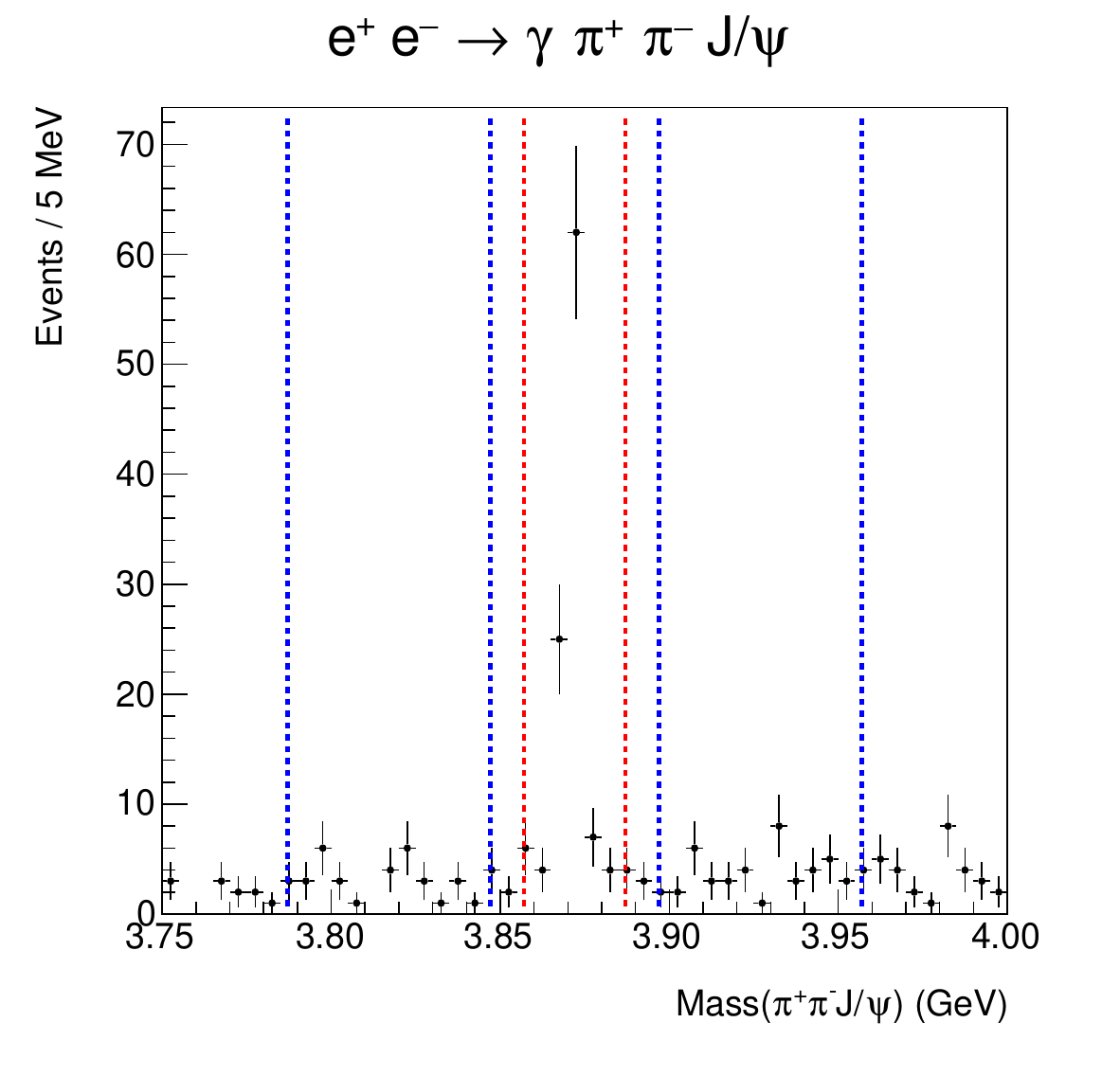}
    \caption{\label{fig:norm_yield} Mass of $\x$ candidates in the normalization channel. The signal region is denoted by red lines and the sideband region by blue lines. A counting method is used to determine the number of $\x$ events in the signal peak. }
\end{figure}

For consistency, this counting method is also used to compute the signal yield in the normalization channel. Here we use the signal and background regions defined in the $\x\to\piz\chicj$ analysis \cite{ryan}: the signal region is defined as the region within $\SI{15}{MeV}$ of the $\x$ mass in the recoil mass spectrum of $\gamma_1$ against the CM, and the background region is defined between $\SI{25}{MeV}$ and $\SI{85}{MeV}$ from the $\x$ mass on both sides in the recoil mass spectrum of $\gamma_1$ against the CM. As for the search channel, the Rolke method is used to compute the uncertainties on the yield. Data measured in the signal and background regions of the normalization channel are plotted in Fig. \ref{fig:norm_yield}.

The measured numbers of events, as well as background ratios, efficiencies, and the resulting signal yields, are listed in Table \ref{tab:yields} for all modes.
These values are used to compute the ratio $R_{\chicj}$, given as
\begin{align}
  \label{eq:bf}
  R_{\chicj} &= \rpipij \nonumber \\&= \frac{N_{\text{search}}}{N_{\text{norm}}} \frac{\epsilon_{\text{norm}}}{\epsilon_{\text{search}}} \frac{1}{\bfpizgg^2\bfchicjtogjpsi},
\end{align}
where $N_\text{search}$ and $N_\text{norm}$ are the measured number of signal events in the search and normalization channels, respectively, and $\epsilon_\text{search}$ and $\epsilon_\text{norm}$ are the efficiencies calculated for the search and normalization channels, respectively. The branching fractions $\bfpizgg$ and $\bfchicjtogjpsi$ are those tabulated by the PDG \cite{pdg}.

\begin{table}[!htbp]
    \caption{\label{tab:yields} Summary of signal yields, efficiencies, and other values used in the calculation of the branching fraction ratios $\frac{\bfpipiot}{\bfnorm}$.}
    \begin{center}
        \begin{tabular}{cccccc}
            \hhline{======}
            Mode & $N_\text{sig}$ & $N_\text{bkg}$ & $\epsilon$ & $\tau$ & $N_\text{search/norm}$ \\
            \hline
            $\xtopipio$ & $\sigregchico$ & $\bkgregchico$ & $\effchico$ & $\tauchico$ & $\yieldchico$ \\
            $\xtopipit$ & $\sigregchict$ & $\bkgregchict$ & $\effchict$ & $\tauchict$ & $\yieldchict$ \\
            \hline
            $\xtopipijpsi$ & $\sigregnorm$ & $\bkgregnorm$ & $\effnorm$ & $\taunorm$ & $\yieldnorm$ \\
            \hhline{======}
        \end{tabular}
    \end{center}
    
\end{table}

\section{Systematic Uncertainties}

A variety of sources of systematic uncertainties on the branching fraction ratio are addressed. In addition to quantities such as the cross-section and luminosity, several uncertainties on the branching fraction $\bfpipiot$ cancel upon computing the ratio. Here we describe these remaining systematic uncertainties on the ratio $R_{\chicj}$, given by Eq. \ref{eq:bf}.

\subsection{Photon detection and charged track efficiencies}

Photons and charged tracks in the BESIII detector were previously measured to have systematic uncertainties of 1\% per photon \cite{photonsys} and charged track \cite{tracksys}. In the ratio, five photons (four from $\piz$ decays and one from the $\chicj$ decay) and two charged pions are uncanceled. Therefore, the photon detection efficiency systematic uncertainty is $5\%$ and the charged track reconstruction efficiency is $2\%$.

\subsection{Input branching fractions}

The remaining branching fractions $\bfpizgg$ and $\bfchicjtogjpsi$ do not cancel in the branching fraction ratio. These branching fractions and their uncertainties are taken directly from the PDG \cite{pdg}.

\subsection{Kinematic fit}

\label{sec:kinfitsys}

The systematic uncertainty due to the kinematic fit is addressed by comparing the effect of a loosened $\chisqdof$ cut to the nominal value on control samples. The process $\controlsearch$ with $\eta\to\g\g\piz\piz$ is used to estimate contributions to the uncertainty from the search channel, while the process $\controlnorm$, with $\psi(3686)\to\pip\pim\jpsi$, is used to estimate contributions from the normalization channel. The efficiency of the $\chisqdof$ selection is computed in signal MC for the search and normalization channels by dividing the number of events with the nominal $\chisqdof$ by the number of events with a loose selection of $\chisqdof<25.0$. The same efficiency is then calculated in data for both channels. The ratio between the search and normalization channels is then computed in MC and in data. The difference between the values of these ratios, $\kinfitsyso\%$ for $\chico$ and $\kinfitsyst\%$ for $\chict$, is taken as the systematic uncertainty.

\subsection{$E/p$ cut}

The systematic uncertainty on the $E/p$ cut is computed in the same way as the kinematic fitting systematic uncertainty in Section \ref{sec:kinfitsys}. The nominal selection criteria in the search and normalization control samples are relaxed to $E/p<0.5$ for both muons and $E/p>0.5$ for either electron. Ratios of efficiencies between the search and normalization channels are calculated in MC and data and the difference between these ratios, $\eopsyso\%$ for $\chico$ and $\eopsyst\%$ for $\chict$, is taken as the systematic uncertainty in the $E/p$ selection.

\subsection{Decay model}

Generating signal MC events containing an $\x$ requires an assumption about the decay model used to describe the production mechanism $\EE\to\g\x$. This results in a systematic uncertainty that we estimate by generating $\EE\to\g\x$ signal MC with four alternative decay models. The nominal model is based on the model for the decay $\psi(3686)\to\g\chico$; the process is assumed to be \textsc{E1} dominant where the photon has the angular distribution $1 - \frac{1}{3}\cos^2\theta$ \cite{p2gc1}. The four variations are: (1)~phase space; (2)~$L=0$, $S=1$; (3)~$L=2$, $S=1$; and (4)~$L=2$, $S=2$. Here, $S$ is the combined spin of the $\gamma$ and the $\x$ and $L$ is the orbital angular momentum between them.

We further introduce two systematic variations on the decay model for $\x\to\rho^0\jpsi$ used for the normalization channel.  These are (1)~$L=2$, $S=1$ and (2)~$L=2$, $S=2$, where $S$ is the combined spin of the $\rho^0$ and $\jpsi$ and $L$ is the orbital angular momentum between them.  These two configurations, alongside the nominal $S$-wave, are the only partial waves allowed for $\x\to\rho^0\jpsi$ by angular momentum conservation.

The greatest difference between efficiency ratios computed with the nominal model and one of the model variations, $\modelsyso\%$ for $\chico$ and $\modelsyst\%$ for $\chict$, is taken to be the systematic uncertainty.

\subsection{Efficiency ratio energy dependence}

In order for the cross section and luminosity to cancel in the ratio given in Eq. \ref{eq:bf}, we assume the reconstruction efficiency at each CM energy used in the calculation is identical. In practice, this assumption is not exactly true, resulting in an additional systematic uncertainty caused by some energy dependence on the efficiency ratio.
To assess this uncertainty, the weighted average of the efficiency ratio over the center-of-mass-energy range is calculated, with the weight of each energy point given by the value of the nominal $\sigma(\EE\to\g\x)$ lineshape. The resulting average is taken as the nominal value. The mass and width parameters of this lineshape are then varied by one sigma in either direction in every possible combination and an average efficiency ratio is calculated for each. The greatest deviation from the nominal value, $\enerdepsyso\%$ for $\chico$ and $\enerdepsyst\%$ for $\chict$, is taken to be the systematic uncertainty.

\subsection{Total systematic uncertainties}
All systematic uncertainties for a particular mode are summed in quadrature to give the total systematic uncertainty, summarized in Table \ref{tab:systematics}.

\begin{table}[!htbp]
    \caption{\label{tab:systematics} Relative systematic uncertainties, given in percent. }
    \begin{center}
        \begin{tabular}{ccc}
            \hhline{===}
            Source & $\chico$ (\%) & $\chict$ (\%)\\
            \hline
            Photon selection & $\phantom{0}\showersyso$ & $\showersyst$ \\
            Tracking & $\phantom{0}\tracksyso$ & $\tracksyst$ \\
            Input branching fractions & $\phantom{0}\inputbfso$ & $\inputbfst$ \\
            Kinematic fit & $\phantom{0}\kinfitsyso$ & $\kinfitsyst$ \\
            $E/p$ & $\phantom{0}\eopsyso$ & $\eopsyst$ \\
            Decay model & $\modelsyso$ & $\modelsyst$ \\
            Energy dependence & $\phantom{0}\enerdepsyso$ & $\enerdepsyst$ \\
            \hline
            Total & $\totalsyso$ & $\totalsyst$ \\
            \hhline{===}
        \end{tabular}
    \end{center}
\end{table}

\section{Upper Limit Calculations}

Due to the low significance of the signal yields for both the $\chico$ and $\chict$ modes, we calculate upper limits on the number of signal events in the search channel using the Rolke method \cite{rolke}, at the $90\%$ confidence level. In addition, the total systematic uncertainties given in Table \ref{tab:systematics}, added in quadrature with the maximum uncertainty on the normalization channel yield, are treated as an uncertainty on the efficiency. These upper limits are then divided by the normalization channel measurement to obtain the upper limit on the ratio. The resulting upper limits on the branching fraction ratios are

\vspace{-\baselineskip}
\begin{align}
\frac{\bfpipio}{\bfnorm} < \ulchico
\end{align}
and

\vspace{-\baselineskip}
\begin{align}
\frac{\bfpipit}{\bfnorm} < \ulchict.
\end{align}
These upper limits on the branching fraction ratios serve to further disfavor the conventional charmonium hypothesis for the $\x$ and are consistent with compact tetraquark and hadronic molecule predictions.

\section{Summary}

Our search for the decays $\xtopipiot$ in BESIII data between $\SI{4.15}{GeV}$ and $\SI{4.30}{GeV}$ did not yield any evidence for these processes. The upper limits of $R_{\chicj}$ at the 90\% confidence level were determined for both channels, with $\frac{\bfpipio}{\bfnorm} < \ulchico$ and $\frac{\bfpipit}{\bfnorm} < \ulchict$. The $\chico$ result clearly disfavors a conventional charmonium interpretation of the $\x$, which is expected to exhibit a branching fraction ratio of $\frac{\bfpipio}{\bfpio}\approx12.5$ \cite{ccbar}. In addition, both results are consistent with models describing $\x$ as a four-quark state \cite{molecular}, although the statistics at the moment is insufficient to validate these predictions.

\acknowledgments

The BESIII Collaboration thanks the staff of BEPCII and the IHEP computing center for their strong support. This work is supported in part by National Key R\&D Program of China under Contracts Nos. 2020YFA0406300, 2020YFA0406400, 2023YFA1606000; National Natural Science Foundation of China (NSFC) under Contracts Nos. 11635010, 11735014, 11835012, 11935015, 11935016, 11935018, 11961141012, 12025502, 12035009, 12035013, 12061131003, 12192260, 12192261, 12192262, 12192263, 12192264, 12192265, 12221005, 12225509, 12235017; the Chinese Academy of Sciences (CAS) Large-Scale Scientific Facility Program; the CAS Center for Excellence in Particle Physics (CCEPP); Joint Large-Scale Scientific Facility Funds of the NSFC and CAS under Contract No. U1832207; CAS Key Research Program of Frontier Sciences under Contracts Nos. QYZDJ-SSW-SLH003, QYZDJ-SSW-SLH040; 100 Talents Program of CAS; The Institute of Nuclear and Particle Physics (INPAC) and Shanghai Key Laboratory for Particle Physics and Cosmology; European Union's Horizon 2020 research and innovation programme under Marie Sklodowska-Curie grant agreement under Contract No. 894790; German Research Foundation DFG under Contracts Nos. 455635585, Collaborative Research Center CRC 1044, FOR5327, GRK 2149; Istituto Nazionale di Fisica Nucleare, Italy; Ministry of Development of Turkey under Contract No. DPT2006K-120470; National Research Foundation of Korea under Contract No. NRF-2022R1A2C1092335; National Science and Technology fund of Mongolia; National Science Research and Innovation Fund (NSRF) via the Program Management Unit for Human Resources \& Institutional Development, Research and Innovation of Thailand under Contract No. B16F640076; Polish National Science Centre under Contract No. 2019/35/O/ST2/02907; The Swedish Research Council; U. S. Department of Energy under Contract No. DE-FG02-05ER41374.

\bibliographystyle{apsrev4-2}
\bibliography{references}

\end{document}